\documentclass[showpacs,aps,superscriptaddress,prl,twocolumn,amsmath]{revtex4}
\usepackage{txfonts}
\usepackage{color}
\usepackage{amssymb}
\usepackage{graphicx}
\usepackage{dcolumn}
\usepackage{bm}
\usepackage{hyperref}
\begin{document}
\preprint{APS/123-QED}
\title{Half-Semiconductor antiferromagnets and Spin-Gapless-Semiconductor antiferromagnets}
\author{Junjie He}
\affiliation{Department of Physics, Xiangtan University, Xiangtan 411105, China}
\author{P. Zhou}
\affiliation{Department of Physics, Xiangtan University, Xiangtan 411105, China}
\author{N. Jiao}
\affiliation{Department of Physics, Xiangtan University, Xiangtan 411105, China}
\author{L. Z. Sun}
\email{lzsun@xtu.edu.cn} \affiliation{Department of Physics, Xiangtan University, Xiangtan 411105, China}
\author{Xiaoshuang Chen}
\affiliation{National Laboratory for Infrared Physics, Shanghai Institute of Technical Physics, Chinese Academy of Sciences, Shanghai, 200083,  China}
\author{Wei Lu}
\affiliation{National Laboratory for Infrared Physics, Shanghai Institute of Technical Physics, Chinese Academy of Sciences, Shanghai, 200083,  China}

\date{\today}
\begin{abstract}
\indent We propose a concept of half-semiconductor antiferromagnets in which both spin-polarized valence and conduction bands belong to the same spin channel with completely compensated spontaneous magnetization. Using density functional theory plus Hubbard U (DFT+U) methods, we find a viable approach to achieve the half-semiconductor antiferromagnets through the transition metal (TM) Fe and Cr codoped boron nitride(BN) sheet. Moreover, spin gapless semiconductor antiferromagnets with zero magnetic moment are also achieved in such systems.\\
\end{abstract}
\pacs{71.20.-b, 71.70.Ej, 73.20.At} \maketitle
\begin{figure}
 \includegraphics[width=3.5in]{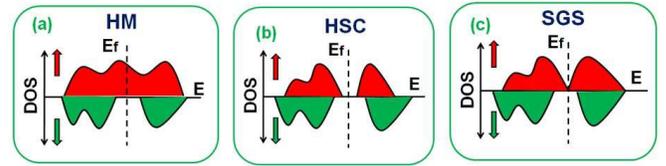}
 \caption{Electronic structure diagrams for (a) Half metal(HM),
(b)Half-semiconductor(HSC), and (c) Spin gapless semiconductor(SGS).}
 \label{fig1}
\end{figure}

\indent The main goal of spintronics is the manipulation of spin degrees of freedom in electronic devices which is  expected  as  the next-generation technology\cite{1}. A tremendous amount of research devoted to design new materials to manipulate spin-polarized current for the practical application in spintronics. Half-metals (HM)\cite{2}, the most important material for spintronics, behave as metallic in one spin channel, but as insulator or semiconductor in the opposite spin channel as shown in Fig.\ref{fig1}(a). The HM have shown potential applications in spintronics as spin filters, detectors, sensors, and a theoretically infinite magnetoresistance\cite{3}. Half-semiconductor (HSC)\cite{4} show semiconductor characteristics with fully spin-polarization in the same spin channel at the valence band maximum (VBM) and the conduction band minimum (CBM) as shown in Fig.\ref{fig1}(b). Such HSC material is a new potential candidate to generate and manipulate spin currents for spintronics. Previous studies indicate that the like-half-semiconductor shows promising application in such as the negative magnetoresistance\cite{5} and spin injection/detection\cite{6}. Spin gapless semiconductors(SGS)\cite{7} with fully spin polarized electrons and holes as shown in Fig.\ref{fig1}(c), has been proposed firstly by Wang\cite{7}. Some remarkable phenomena have been observed experimentally in the SGS materials containing colossal electroresistance, giant magnetoresistance\cite{8}, and crossover of magnetoresistance\cite{9}. Most intriguingly, Mn$_2$CoAl with a SGS has been successfully synthesized with a Curie temperature of 720 K and exhibits nearly temperature-independent conductivity and carrier concentration\cite{10}. Moreover, the HSC and SGS are similar to that of HM show quantized net magnetic per unit cell due to the gap of their majority and minority channels.\\
\indent In general, the HM, HSC and SGS show spin polarization and behave as ferromagnets. However, van Leuken and de Groot\cite{11} indicated the coexistence of  HM and antiferromagnetic (fully compensated magnetic moment) as half metallic antiferromagnets (HMAF)\cite{11,12}. The new materials possessing the half-metallic characteristics but without any net magnetization have been theoretically predicted in V$_7$MnFe$_8$Sb$_7$In alloys\cite{11}, perovskite oxides\cite{13,14}, Mott insulator NiO\cite{15}, monolayer superlattice\cite{16}, and diluted magnetic semiconductors\cite{17}. HMAF is potentially used as novel spin polarized  STM tips\cite{12}, a new spin injection devices and single spin superconductivity\cite{13} due to its unique properties. Thus far, the most interesting issues is that: whether the antiferromagnetic (fully compensated magnetic moment) HSC and SGS without any net magnetization can be achieved. Such materials might be show significant potential application in spintronics due to their semiconducting characteristics. These class of materials are named as half-semiconductor antiferromagnets (HSCAF) and spin gapless semiconductor antiferromagnet (SGSAF). The concept of SGSAF has been proposed in the report of Wang\cite{7}, however, it has not been achieved in real materials yet. In present letter, we find a viable approach to achieve the HSCAF and SGSAF through the transition metal (TM) Fe and Cr codoping boron nitride(BN) sheet. Such a full spin polarized material with compensated zero magnetic moment not only provide two promising candidates to spintronics, but also enrich the concept of the magnetic order.\\
\begin{figure}
 \includegraphics[width=3.0in]{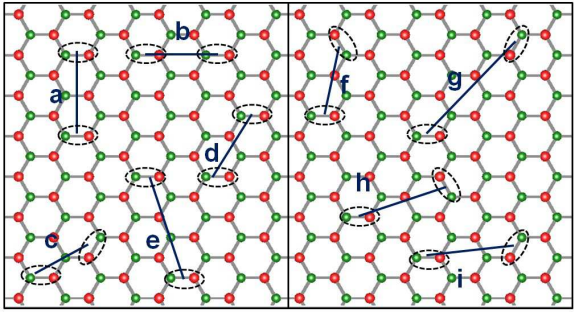}\\
 \caption{Schematic of the typical Fe and Cr codoping configurations on the B-N divacancy of BN. Even though each supercell has 110 atoms, only the relevant portions of the supercells are shown in the figure for clarity.}
 \label{fig2}
\end{figure}
\indent Two dimensional (2D) monolayer materials, such as graphene\cite{18,19}, have shown the unique properties and tremendous possible applications in nanoscale devices. Previous works\cite{20,21,22,23,24,25} have shown that TM impurity profoundly influences the electronic and magnetic properties of graphene. However, the zero band gap of graphene lead to the TM d states strongly couple with its conduction or valence bands. It is difficult to obtain the electronic structures like HSC and HM by TM doping. Recently, some wide gap semiconductor, such as monolayer BN, AlN, GaN, and ZnO, have been theoretically predicted\cite{26,27} and successfully synthesized in experiments\cite{28,29} as promising materials for spintronics. When TM doped in the wide band gap semiconductor,  the electronic structures of the systems are primarily determined by the distribution of TM d states around the Fermi level because the energies of TM 3d orbitals can exist inside the band gap, by which some novel magnetic order can be achieved, for example,  TM doped BN sheet shows the HSC, HM and SGS magnetic characteristics\cite{30,31,32}. Here we take  BN sheet as typical template to explore the possibility of the achievement of HSCAF and SGSAF with TM pairs codoping technique. We find that using the TM codoping technique to obtained the HSCAF and SGSAF three conditions are required: (i) the same spin states for the two TM; (ii) the coupling between the TMs and the host semiconductor maintains antiferromagnetic order of the TMs; (iii) the host semiconductor is a wide band gap semiconductor. Previous work\cite{31,33} have indicated that when the single Fe and Cr doped on the B-N divacancy (DV) of BN sheet (here the BN sheet with DV is named as BN-DV for short) show not only extremely stability but also the same high spin state with 4$\mu_B$ magnetic moments. It suggest that if Fe and Cr maintain a antiferromagnetic coupling, the system might be a possible candidate for HSCAF or SGSAF.\\
 \begin{figure}
 \includegraphics[width=3.5in]{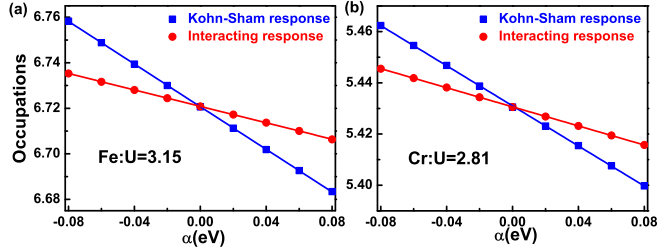}\\
 \caption{The interacting ($\chi$) and Kohn-Sham response ($\chi_0$) functions of single Fe and Cr doped on BN-DV system.}
 \label{fig3}
\end{figure}
\indent Previous studies\cite{25,42} indicated that as for TM adsorbed graphene, graphyne, and graphdiyne, the magnetic states is sensitive to Hubbard U. Therefore, to account for the correlation energy of the strongly localized 3d orbital of TM, Hubbard U correction is employed (DFT+U ) with the generalized gradient approximation of Perdew Burke Ernzerhof (PBE)\cite{34} implemented in Vienna ab initio simulation package (VASP)\cite{35,36}. We determined the parameter U for Fe and Cr doped on DV of BN by the linear response approach introduced by Cococcioni et al.\cite{37}. In this approach, the interacting ($\chi$) and Kohn-Sham response($\chi_0$) functions of the system with respect to localized perturbations are firstly calculated. And then the parameter U can be obtained from the formula: $U=\chi^{-1}-\chi^{-1}_0$. The calculated effective U values for Fe and Cr are shown in Fig.\ref{fig3}. In our calculations, a supercell contained 110 atoms is adopted to be the calculated model, where a vacuum space of $15\AA$ perpendiculars to the BN plane is chosen. A plane-wave kinetic energy cutoff of 500 eV is employed. The Brillouin zone (BZ) is sampled using 5x5x1 and 9x9x1 Gamma-centered Monkhorst-Pack grids for the calculations of relaxation and electronic structures, respectively. The criteria of energy and atom force convergence are set to $10^{-5}$ eV/unit cell and 0.015 eV/$\AA$, respectively. \\
\begin{table*}
\caption{The calculated results of Fe and Cr codoped BN-DV system under nine typical configurations. The M.O. represents the magnetic order between the Fe and Cr atom. The $\Delta$ E represents the relative energy for different Fe and Cr codoped configurations. E$_{ex}$ is the energy difference defined as $E_{ex}$=$E_{FM}$-$E_{AFM}$, where $E_{FM}$ and $E_{AFM}$ are the total energies of Fe and Cr codoped BN-DV system under ferromagnetic and antiferromagnetic states, respectively. The $M_{Cr}$, $M_{Fe}$, $M_{AF}$, and $M_{FM}$ are the local magnetic moment of TM atom and total magnetic moment under FM and AF states per unit cell. The 'ES' denote the electronic structure of codoped system. $E_{gap}$ defined the difference between HOMO and LUMO as $E_{gap}$=$E_{HOMO}-E_{LUMO}$. The D is the distance of Fe and Cr for different configurations.}\label{tab1}
\begin{ruledtabular}
\begin{tabular}{ccccccccccc}
\hline
 Con.&M.O.&$\Delta$ E&$E_{ex}(meV)$&$M_{Cr}$($\mu_B$)&$M_{Fe}$($\mu_B$)&$M_{AF}$($\mu_B$)&$M_{FM}$($\mu_B$)&ES&$E_{gap}$(eV)&D($\AA$) \\
\hline
a        &AF &0.738 &11  &3.686 &-3.466 &0 &8&HSCAF&0.38 &5.12 \\
b        &AF &0.407 &18  &3.724 &-3.503 &0 &8&HSCAF&1.06 &4.30  \\
c        &AF &0.674 &107 &3.583 &-3.375 &0 &8&HSCAF&0.65 &4.29  \\
d        &AF &0     &32  &3.362 &-3.179 &0 &8&SGSAF&0    &4.24  \\
e        &AF &0.681 &1.3 &3.692 &-3.482 &0 &8&HSCAF&0.32 &6.67  \\
f        &AF &0.512 &67  &3.647 &-3.403 &0 &8&HSCAF&0.36 &4.84  \\
g        &AF &0.783 &0.7 &3.686 &-3.459 &0 &8&HSCAF&0.68 &8.31 \\
h        &AF &0.677 &3.6 &3.677 &-3.474 &0 &8&HSCAF&0.66 &6.00\\
i        &AF &0.849 &3.1 &3.661 &-3.460 &0 &8&HSCAF&0.58 &5.88 \\
\end{tabular}
\end{ruledtabular}
\end{table*}
 \begin{figure}
 \includegraphics[width=3.0in]{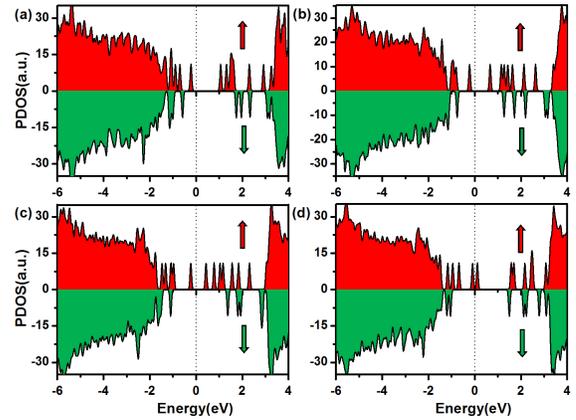}\\
 \caption{Total density of states (DOS) of typical configuration \emph{b} (a), \emph{c} (b), \emph{f} (c), and \emph{d} (e). The solid (red) and dash (blue) represent the majority and minority spin channels, respectively. The Fermi level is set to zero.}
 \label{fig4}
\end{figure}
\indent Huang et al.\cite{33} proposed that because of the energy of TM 3d orbitals are roughly located within the wide energy gap of BN, the TM could be as TM$^{2+}$ ion after interact with the in gap defect states of the double-acceptor DV. We performed the DFT+U calculations to study single Cr and Fe doeped BN-DV. The impurity complex Fe/Cr-DV system is found to be magnetic with the magnetic moment 4 $\mu_B$ per unit cell. The results suggest the Cr and Fe behave as Cr$^{2+}$ and Fe$^{2+}$ ions on the DV of  BN with high spin (HS) states with the $d^{4\uparrow}d^{0\downarrow}$ and $d^{5\uparrow}d^{1\downarrow}$ spin configuration, respectively. This conclusion is similar to the report of other groups\cite{31,33}. We use Cr and Fe pairs to codope BN sheet and each of them fill a DV separately. We only consider non-merged DV cases in which Fe and Cr atom maintain a indirect interaction without direct Fe-Cr bonds. All possible configurations are considered and schematically summarized in Fig.\ref{fig2}. Table \ref{tab1} shows the relative energy with respect to the most stable configuration \emph{d} as shown in Fig.\ref{fig1} that Fe and Cr atoms doped into two next nearest neighbor (NN) nonmerging DV of BN with the distance of 4.24 $\AA$. The sequence of relative stability of typical TM codoped BN-DV systems is \emph{d}$>$\emph{b}$>$\emph{f}$>$\emph{c}$>$\emph{h}$>$\emph{e}$>$\emph{a}$>$\emph{g}$>$\emph{i}. In general, the relative energy is depend on the distance of the two TMs in nonmerged DV. For instance, the configuration  \emph{e}, \emph{g} and \emph{h} show the higher relative energy with larger distance between TMs.\\
\indent For all the Fe and Cr codoped BN-DV systems,  the magnetic moments are 8 $\mu_B$ and 0 $\mu_B$ for the system under FM and AFM coupling states, respectively.  The results of local magnetic moments indicate that both Fe and Cr show the HS states with  $d^{5\uparrow}d^{1\downarrow}$ and $d^{4\uparrow}d^{0\downarrow}$ spin configuration, respectively; they behave as Fe$^{2+}$ and Cr$^{2+}$ ions pairs in BN-DV system. It is similar to the case of single Fe and Cr doped BN-DV systems. Table \ref{tab1} lists the the energy differences ($\Delta E$=$E_{FM}$-$E_{AF}$) between FM and AFM states for all codoped configurations, as well as the spin magnetic moment of the Fe$^{2+}$ and Cr$^{2+}$ ions. A number of very interesting properties can be observed in Tab.\ref{tab1}, with the most intriguing being the AFM coupling with zero net magnetic moment for all configurations. Depending on the orientation and the distance between Fe$^{2+}$ and Cr$^{2+}$ ions, the results demonstrate considerable difference of the exchange energy. Such phenomenon originates from the dependence of the coupling between Fe and Cr on the mediation of B/N atoms. According to previous reports, the Co dimer embedded the DV of graphene nanoribbons (GNRs) lead to a magnetic anisotropy of FM and AF coupling because the spin polarization induced by both Co and ribbon edge.\cite{38} Unlike the magnetic coupling anisotropy of Co dimer in DV GNRs,\cite{38} each Fe and Cr codoped BN-DV configuration shows antiferromagnetic coupling. Moreover, Fe and Cr also induce the magnetic moments of their neighboring B and N atoms, but the total magnetic moment of the systems is zero indicating a completely compensated antiferromagnets. The above results indicate that the Fe and Cr codoped BN-DV system can achieve the completely compensated antiferromagnet is due to both TMs show same spin states in the systems.\\
\indent From the results of the electronic structures of Fe-Cr codoped BN-DV sheet, the density of states (DOS) of some typical systems are shown in Fig.\ref{fig4}, we find that 100$\%$ spin polarization is shown in the HOMO and LUMO for the same majority spin channel with a vanishing net magnetization. Therefore, Fe-Cr codoped BN-DV systems achieve the magnetic order of HSCAF. For example, the configurations \emph{b}, \emph{c}, and \emph{f} as shown Fig.\ref{fig4}(a), (b), and (c) demonstrate that both the majority and minority spin channels are semiconductor and the HOMO and the LUMO states belong to the same majority spin channel with the energy gap (the difference between HOMO and LUMO states, $E_{gap}$=$E_{HOMO}-E_{LUMO}$) of 1.06 eV, 0.65 eV and 0.36 eV, respectively. The three typical systems behave as HSCAF. However, as for the configuration \emph{d} as shown in Fig.\ref{fig4} (d), the HOMO and LUMO of the majority spin touch each other at the Fermi level, whereas the minority spin channel shows a insulator gap. The results of energy band structures indicate that the majority spin channel has a small gap about 0.09 eV. According to the definition of SGS by Wang et al.\cite{7} that the term "gapless" is used for an narrow energy gap that is approximately smaller than 0.1 eV. Therefor the most stable configuration \emph{d} realizes the SGSAF proposed by Wang\cite{7}. For the other configurations, our results indicate that similar electronic structures with HSCAF can be obtained as listed in Tab.\ref{tab1}. Overall, our results reveal that HSCAF and SGSAF properties can be achieved via Fe and Cr codoped BN-DV systems.\\
 \begin{figure}
 \includegraphics[width=3in]{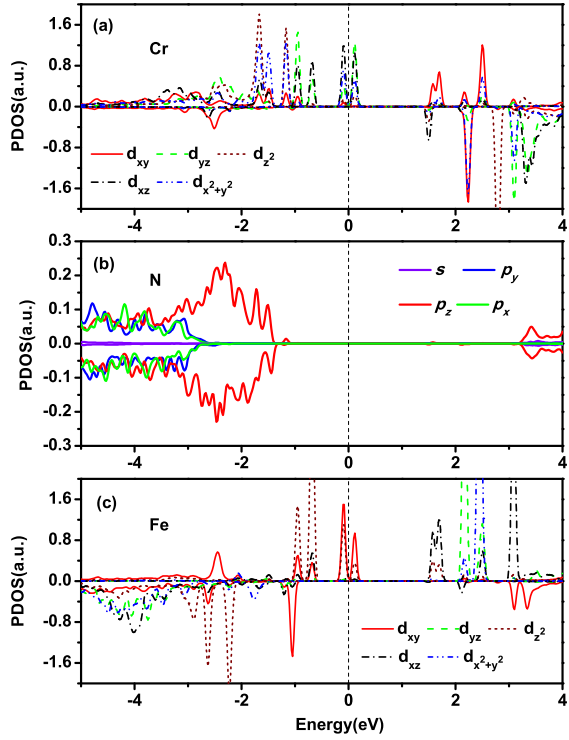}\\
 \caption{Partial density of states (PDOS) of the most stable configuration \emph{d}. The PDOS of the d orbitals of TMs and their nearest neighbor (NN) N atom are also presented. The Positive and negative values denote majority and minority channels, respectively. The Fermi level is set to zero.}
 \label{fig5}
\end{figure}
\indent To explain the exotic electronic structures of Fe-Cr codpoed BN-DV systems, it is needed to thoroughly study the bonding mechanism along with the AF coupling between Cr$^{2+}$ and Fe$^{2+}$. We take the the most stable configuration \emph{d} as an example to illustrate the origin of antiferromagnetic coupling due to the similar characteristics of the electronic structures of the codoped systems. From the PDOS of the configuration \emph{d} as shown in Fig.\ref{fig5}, the majority d states of Cr atom mainly lie below the Fermi level within the [-2.1 eV, 0 eV] energy window, whereas some unoccupied majority d state peaks locate around 0.1 eV, 1.6 eV and 2.5 eV, respectively. Most of the d states of Cr locate in the gap of the electronic states derived from the host template, where only slightly overlap between $p_z$ orbitals of N atom and $d_{z^2}$ and $d_{x^2+y^2}$ orbitals of the Cr atom within [-1.9 eV, -1.3 eV] energy window. Moreover, the d states of Cr below the Fermi level are predominantly in the majority channel; and they are close to Cr $d^{4\uparrow}d^{0\downarrow}$ spin configuration according to the integration of the PDOS. As for the case of Fe atom, the minority d states are almost all occupied, moreover, the majority d states also show occupation especially within the energy window [-1.0eV, 0.0eV]. The occupied majority d states almost exist inside the band gap of the host material as that of Cr. Furthermore, the similar majority peaks for both Fe and Cr around the Fermi level indicate their coupling effects. Compare with the Cr atom, the minority d states of Fe have a overlap with not only the $p_z$ states but also $p_x$ and $p_y$ states of N atoms; moreover, the energy window of the predominant minority channel is lower than the predominant majority channel of Cr. According to the integration of the PDOS of Fe-d, the Fe atomic spin configuration is close to $d^{1\uparrow}d^{5\downarrow}$. \\
 \begin{figure}
 \includegraphics[width=3in]{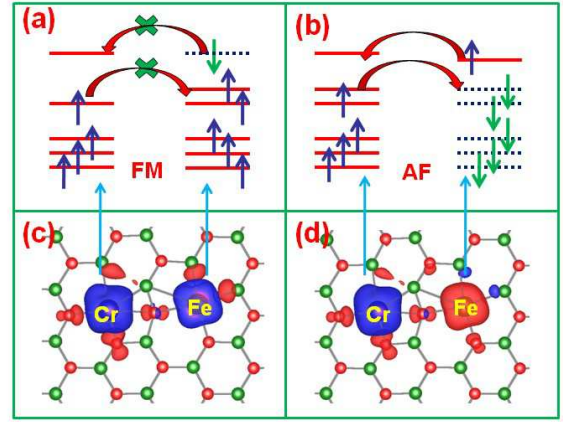}\\
 \caption{The d orbitals of codoped Cr$^{2+}$ and Fe$^{2+}$ ions under (a) FM  and (b) AF coupling. The blue and green arrows indicate spin up and spin down electron, respectively, and the dotted arrows and cross indicate permission or forbidden of electron hopping. (c) and (d) are the spin-polarized charge density (SCD) distribution for configuration \emph{d} under FM and  AF coupling, respectively. The dark (red) and light (green) isosurfaces denote the spin-up and spin-down charge density, respectively. The isosurface of SCD is 0.005 $e/\AA^3$ }
 \label{fig6}
\end{figure}
\indent In present work, we take a simplified model to analyze the magnetic coupling mechanism between Cr$^{2+}$ and Fe$^{2+}$, the typical results of configuration \emph{d} are shown in Fig.\ref{fig6}. Fig.\ref{fig6}(a) and (b) illustrate the energy level of d orbitals of Fe$^{2+}$($d^{1\uparrow}d^{5\downarrow}$) and Cr$^{2+}$($d^{4\uparrow}d^{0\downarrow}$) and their possible electron hopping under FM and AFM coupling, respectively. Fig.\ref{fig6}(c) and (d) display the spin charge density (SCD) for the system under FM and AF coupling, respectively. As for FM coupling, Cr$^{2+}$ and Fe$^{2+}$ have tow possible virtual hopping features for the frontier orbitals including the Cr$^{2+}$ transfer one spin up electron to spin up electron half-occupied orbital of Fe$^{2+}$; or Fe$^{2+}$ transfer one electron to unoccupied orbital of Cr$^{2+}$. However, the exchange interaction with transfer of one spin up electron from Cr$^{2+}$ to one spin up electron half-occupied orbital of Fe$^{2+}$ require extra \emph{Hund}' intra-atom exchange energy between spin-up and spin down states. As for the case of AF coupling, the exchange virtual hopping ways between Fe$^{2+}$ and Cr$^{2+}$ are unblocked, the AF coupling between Fe$^{2+}$ and Cr$^{2+}$ does not need any additional \emph{Hund}' coupling energy. Therefore,  the AF coupling between Fe$^{2+}$ and Cr$^{2+}$ is more favorable; the systems behave as antiferromagnet. Based on the analysis above, the magnetic exchange interaction of Fe$^{2+}$ and Cr$^{2+}$ follows the double exchange mechanism as that of diluted magnetic semiconductor and half-metallic diluted antiferromagnetic semiconductor\cite{17}.\\
\indent With the perfect antiferromagnetic coupling, the magnetic moment of Cr$^{2+}$ and Fe$^{2+}$ completely compensate each other resulting in zero magnetic moment of the Fe-Cr codoped BN-DV systems. However, the exchange splitting of Fe and Cr is different due to their different coupling nature with NN N and B atoms. As shown in Fig.\ref{fig5}(a), the majority spin channel of Cr shift towards lower energy with about four electron occupation, while its unoccupied minority spin channel shift toward to higher energy. However, the exchange splitting of Fe$^{2+}$ as shown in Fig.\ref{fig5}(c) produces its majority spin channel to higher energy with about one electron occupation and its minority spin channel to lower energy with five electron occupations. Consequently, the spin channel of the HOMO and LUMO of the codoped systems is the same majority one with 100\% spin-polarization for both VBM and CBM due to the inverse exchange splitting for Cr$^{2+}$ and Fe$^{2+}$. \\
\indent In conclusion, we propose the half-semiconductor antiferromagnets (HSCAF) which shows fully spin-polarized valence and conduction bands with the same spin channel and vanish spontaneous magnetization.  Based on the first-principles method, we design a viable approach to achieve the HSCAF through the transition metal (TM) Fe and Cr codoped on the BN-DV. Moreover, the spin gapless semiconductor antiferromagnet proposed previously is also achieved in real material. Such a full spin polarized semiconductors with compensated zero magnetic moment not only provide promising candidates to spintronics, but also enrich the concept of the magnetic order.\\
\indent For TM adsorbed BN, TM adatoms is energetically favorable for TM cluster on BN sheet due to the relatively small diffusion barriers and weak binding energy\cite{30,32}. However, the adsorption of TM atom on the BN vacancy in BN sheet shows higher binding energy.\cite{31,33} In experiments, the vacancy can be created using electron beam\cite{39,40}. With the help of the accurate manipulation of STM tip, such two new magnetic orders are hoping to be realized in experiments. Moreover, to our knowledge, previous predictions of the HMAF materials are all based on bulk materials. As pointed out by Hu et al.\cite{41} that the complicated lattice structure of the bulk HMAF materials easily accompany with the lattice distortions, which would suppress the bulk HMAF property. Furthermore, the spin-orbital coupling (SOC) may also destroy the fully compensated magnetization. For the above reasons, there are not clear experimental evidence of the existence of the HMAF in bulk material until now. With extremely weak SOC and simple monolayer crystal structure, the Fe and Cr codoped BN-DV systems prior to previous predicted bulk materials are promising candidates to realize the exotic magnetic orders of HMAF,  HSCAF and SGSAF in experiments.\\
\begin{acknowledgements}
This work is supported by the Program for New Century Excellent Talents in University (Grant No. NCET-10-0169), the Hunan Provincial Innovation Foundation for Postgraduate (Grant No. CX2010B250), the National Natural Science Foundation of China (10990104), and the computational support from Shanghai Super-computer Center.
\end{acknowledgements}

\end{document}